\documentclass[aip,11pt,reprint]{revtex4-1} 

\usepackage{amsmath} 
\usepackage{amsfonts}
\usepackage{graphicx}
\usepackage{float}
\usepackage{leftindex}
\usepackage{gensymb}

\begin{document}

\title{Why is there no Poisson spot in a solar eclipse?}

\author{Jami J. Kinnunen}
\affiliation{Department of Applied Physics, Aalto University School of Science, FI-00076 Aalto, Finland}

\begin{abstract}
The Poisson spot is a fascinating lecture demonstration. Its simple explanation can lead to further questions, not only the one posed in the title, but also questions such as why the simple model that considers only light passing just outside the spherical object is successful. The Huygens-Fresnel diffraction model is applied to answer these questions.
\end{abstract}

\maketitle 

\section{Introduction} 

The Poisson spot is an optical phenomenon in which a bright spot of light is seen at the center of a shadow caused by a spherical object. 
It is a simple demonstration to do and works even with white light, but the appearance of the bright spot is an intriguing and surprising effect~\cite{Gluck2010}. Indeed, shadows of spherical objects are quite common, but no Poisson spots are observed in the shadow of various balls or even the Moon. Why?

Huygens' secondary wavelets are often used as an intuitive explanation in undergraduate wave mechanics~\cite{Young}.
The picture is particularly useful for explaining diffraction and interference effects, but it is generally used only for the simplest of cases, such as the far-field interference pattern in double slit experiment. For the Poisson spot, a similar 
simple model can be used, but very quickly one ends up with questions that go beyond the model, such as when considering the (lack of a) Poisson spot in the Moon's shadow.

However, Fresnel's formulation of Huygens' ideas in the Huygens-Fresnel diffraction formula provides a theoretical framework in which Huygens' secondary wavelet picture can be extended to more interesting cases. The theory provides a way to derive simple models in an understandable and transparent way, but also leads to integral equations that can be easily solved by numerical tools.

Physics phenomena that can be demonstrated in lectures\cite{Taylor}, for which simple models can provide intuitive explanations but for which even the more complete and rigorous explanations can be presented understandably are always valuable course material. I have found the Poisson spot in particular to tick all these boxes. Moreover, its counterintuitive nature that can be seen by bare eye creates discussions and feeds the imagination. It has been realized also with (ultra)sound~\cite{Hitachi2010} and molecular beams~\cite{Reisinger2009}, and its potential applications range from astronomy to optical tweezers~\cite{Emile2024}.
 
Below I will describe the Poisson spot experiment and formulate a simple model that allows order of magnitude analysis of the interference maximum. I will then use the Huygens-Fresnel diffraction formula for solving the problem in a more rigorous fashion and confirm the predictions and the underlying assumptions of the simple model.

\section{Poisson spot}

\begin{figure}[h]
    \centering
    \includegraphics[width=0.95\columnwidth]{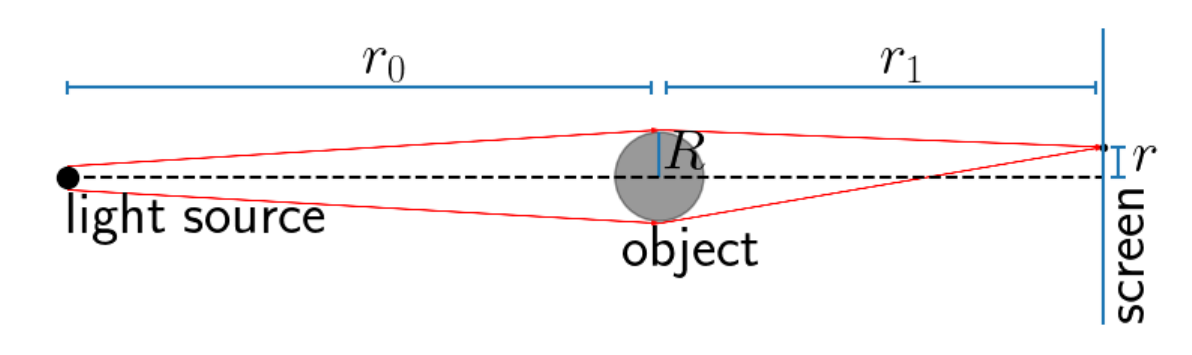}
    \caption{The Poisson spot diffraction experiment involves a source of light with wavelength $\lambda$, an opaque spherical object of radius $R$, and a screen. Light from the source takes all paths around the object, resulting in an interference pattern at the screen. The Poisson spot is an intensity maximum at the center of the shadow ($r=0$) due to constructive interference for the light paths from any edge of the cross-section of the object.}
    \label{fig:setup}
\end{figure}

The history and physics of the Poisson spot, also called the Arago spot or Fresnel spot, is well known~\cite{kelly,harvey}. 
For a perfect opaque sphere blocking a beam of light, the point right at the center of the shadow of the object is at equal distance from the points at the rim of the cross section of the sphere. Thus the waves going around the object interfere constructively, resulting in an intensity maximum. The setup is shown in Figure~\ref{fig:setup}, which also introduces the symbols that will be used in the analysis.

While this intuitive explanation is sufficient for explaining that a bright spot might be observed at the center of the shadow, it raises additional questions such as:
\begin{itemize}
\item how the size of the Poisson spot depends on the wavelength of light, the size of the object and the distance to the screen?
\item how perfectly spherical the object needs to be?
\item what does one mean by 'waves going around the object'?
\item why don't we see Poisson spots in everyday life?
\end{itemize}

\section{Order of magnitude analysis}
\label{sec:orders}

The simple model in which only diffracting light from a narrow ring right around the rim of the object is considered allows one to answer a number of questions presented above. 
The model can be viewed as a cylindrically symmetric extension of the double slit experiment, and many of the key results can be obtained from the model. 

\begin{figure}[h]
    \centering
    \includegraphics[width=0.4\columnwidth]{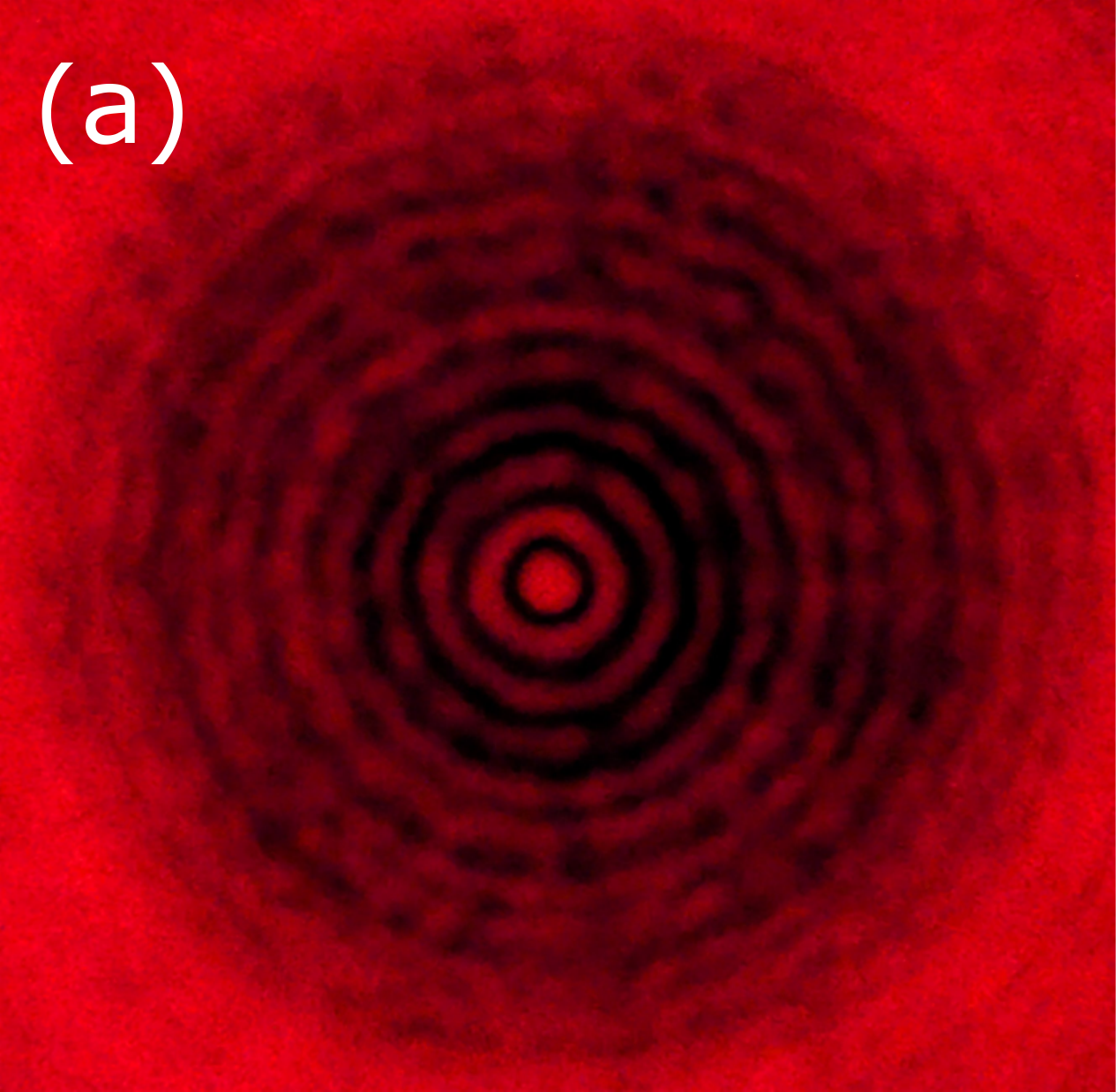}
    \includegraphics[width=0.4\columnwidth]{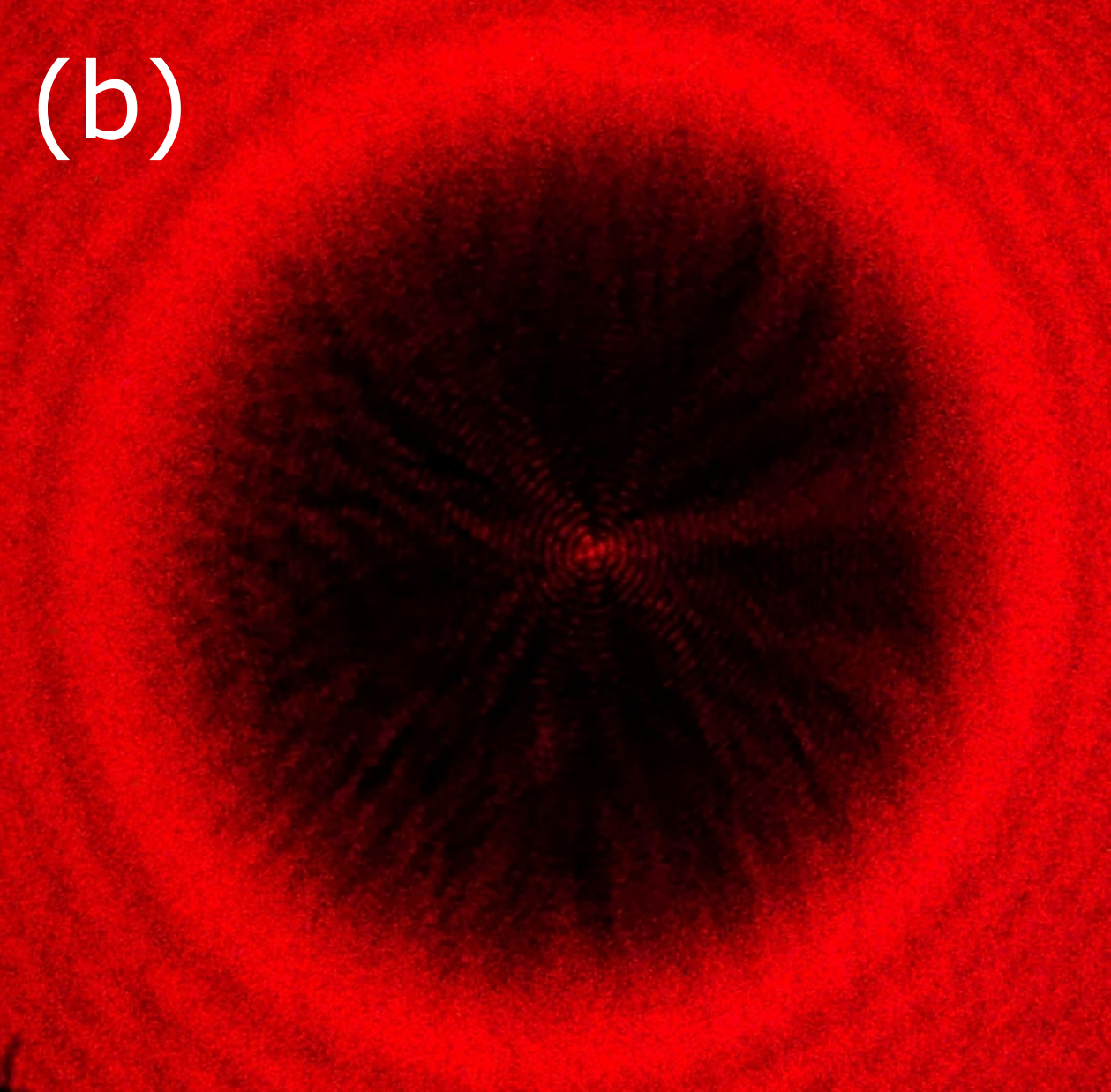}
    \caption{Diffraction patterns with the Poisson spot using a red 630 nm laser and a circular masks of diameter (a) 1 mm and (b) 4 mm. The distance to the screen was 10 meters and the object was positioned in a slowly diverging beam so that the overall shadow at the screen was roughly the same size (diameter 10 cm) for both cases. For the smaller object the diffraction pattern shows a number of interference fringes in addition to the bright central maximum of a four-millimeter wide Poisson spot. In the case of a larger object, the overall shadow is more complete, but the bright Poisson spot is clearly visible although much smaller in size with diameter of roughly one millimeter.} 
    \label{fig:demo}
\end{figure}

A length scale for the radius of the Poisson spot $r_\mathrm{spot}$ can be obtained by assuming that the length difference to two points at the rim of the object equals at most half the wavelength of the light, which would correspond to destructive interference at the screen.
The condition is thus
\begin{equation}
\frac{\lambda}{2} \gtrsim \sqrt{(R+r_\mathrm{spot})^2 + r_1^2} - \sqrt{(R-r_\mathrm{spot})^2 + r_1^2},
\end{equation}
which can be linearized for large $r_1$ to yield
\begin{equation}
\lambda \gtrsim \frac{(R+r_\mathrm{spot})^2}{r_1} - \frac{(R-r_\mathrm{spot})^2}{r_1} = \frac{4Rr_\mathrm{spot}}{r_1}.
\end{equation}
Thus the size of the spot scales as
\begin{equation}
r_\mathrm{spot} \lesssim \frac{\lambda r_1}{4R}.
\label{eq:spotsize}
\end{equation}
Not surprisingly, the size of the spot increases linearly with distance $r_1$, but somewhat more surprisingly it is smaller for larger objects, scaling inversely with the radius $R$. For a typical lecture demonstration setup ($\lambda = 630\,\mathrm{nm}$, $r_1 = 10\,\mathrm{m}$, $R = 0.5\,\mathrm{mm}$) Eq.~\eqref{eq:spotsize} gives $r_\mathrm{spot}= 3\,\mathrm{mm}$, in reasonable agreement with the observed Poisson spot diameter of four millimeters (Fig.~\ref{fig:demo}a).With a larger spherical object, the Poisson spot becomes very sharp and even more pronounced as the overall shadow becomes darker and better defined (Fig.~\ref{fig:demo}b). 
However, as the spot size decreases with the radius of the object, eventually it becomes impossible to observe. With my lecture demonstration setup, a one-centimeter diameter steel bearing is already too large to produce a Poisson spot visible to the unaided eye.

If using a white light point source, the different wavelengths produce spots with different widths, resulting in a white center with a reddish edge. This is quite similar to the case of a double slit experiment using white light, which also has a white spot at the center, since all wavelengths interfere constructively at the center, but the rest of the spots are increasingly colourful.

A similar simple geometric model can be used for calculating the required smoothness of the object.
Assuming a deviation of at most $\delta$ from the perfect circle of radius $R$, the length difference to the center of the shadow may not exceed half the wavelength of the light in order for constructive interference to take place. The condition is
\begin{equation}
\frac{\lambda}{2} \gtrsim \sqrt{(R+ \delta)^2 + r_1^2} - \sqrt{R^2 + r_1^2},
\end{equation}
which can again be solved by linearizing for large $r_1$ to yield
\begin{equation}
\lambda \gtrsim \frac{(R+\delta)^2}{r_1} - \frac{R^2}{r_1} = \frac{4R\delta+\delta^2}{r_1}.
\end{equation}
Assuming that $R \gg \delta$, this can be further simplified to yield
\begin{equation}
\delta \lesssim \frac{\lambda r_1}{4R}.
\label{eq:surfacerough}
\end{equation}
Interestingly this is the same as the size of the spot calculated above. Also the surface needs to be increasingly smooth for larger objects, as the deviation $\delta$ scales inversely with radius $R$. This is quite surprising, as it means that the relative smoothness $\delta/R$ is not the relevant quantity here. For the above lecture demonstration setup,
Eq.~\eqref{eq:surfacerough} yields $\delta = 3\,\mathrm{mm}$, which is already so large that the assumption $R \gg \delta$ fails. In contrast, for the case of the Moon's shadow in solar eclipse, one can estimate using visible light $\lambda \approx 500\,\mathrm{nm}$, Earth-Moon distance $r_1 \approx 4 \cdot 10^5\,\mathrm{km}$ and the radius of the Moon $R \approx 2000\,\mathrm{km}$, the required surface roughness may not exceed $\delta \leq 100\,\mathrm{\mu m}$. Clearly the Moon, or any planetary object, cannot be sufficiently smooth to produce a visible Poisson spot, answering the question in the title of this article.
For a more everyday object, one can consider a soccer ball ($R = 10\,\mathrm{cm}$), which for visible light and distance of one meter would yield the condition of $\delta \leq 5\,\mathrm{\mu m}$. This, too, is clearly beyond the scope of everyday objects.

These simple models can be understood as generalizations of the double slit experiments, since only the path length differences of two points are considered at a time. 
However, we should now return to the question of whether it is justified to consider only a narrow ring of secondary wavelets passing around the object.

\section{Huygens-Fresnel diffraction integral}

The Huygens-Fresnel principle states that each point in a wavefront is a source of a spherical secondary wavelet, and the interference of all these propagating wavelets provides the propagation of the whole wavefront~\cite{born_and_wolf_1999}. 
Neglecting temporal variations of the phase, i.e. considering a snapshot of the diffracting field, a monochromatic point source produces a displacement field at distance $r$ that is
\begin{equation}
U_0(r) = A_0\frac{e^{ikr}}{r},
\end{equation}
where $k=2\pi/\lambda$ is the wave number of the field. The displacement field $U({\bf r})$ describes the amplitude and phase of the modulation of a generic wave medium: it can describe scalar fields, such as progating sound (pressure) waves, or vector fields, such as electric field of a radiating electromagnetic (light) field. While the real part of the displacement field corresponds to the physically measurable quantity, the amplitude $|U({\bf r})|^2$ will give the time averaged intensity, which is the quantity of interest for Poisson spot.

For an extended source, the field at point $P$ is obtained by integrating over the source surface $\mathcal{S}$ to give
\begin{equation}
U(P) = \int_\mathcal{S} A(S)\frac{e^{iks}}{s} dS,
\end{equation}
where $dS$ is an infinitesimal surface element of the source, $s$ is the distance from the surface element to the point $P$, and the prefactor $A(S)$ is the displacement field at the surface element. Notice that this integral formula neglects the inclination factor introduced by Fresnel and later improved upon by Kirchoff~\cite{born_and_wolf_1999}. The inclination factor essentially directs the secondary wavelets in the forward direction. However, when the distances from the light source to the object and from the object to the screen are much larger than the size of the object, the changes in the direction are very small, making the inclination factor so close to unity that it can be neglected.

\begin{figure}[h]
    \centering
    \includegraphics[width=0.95\columnwidth]{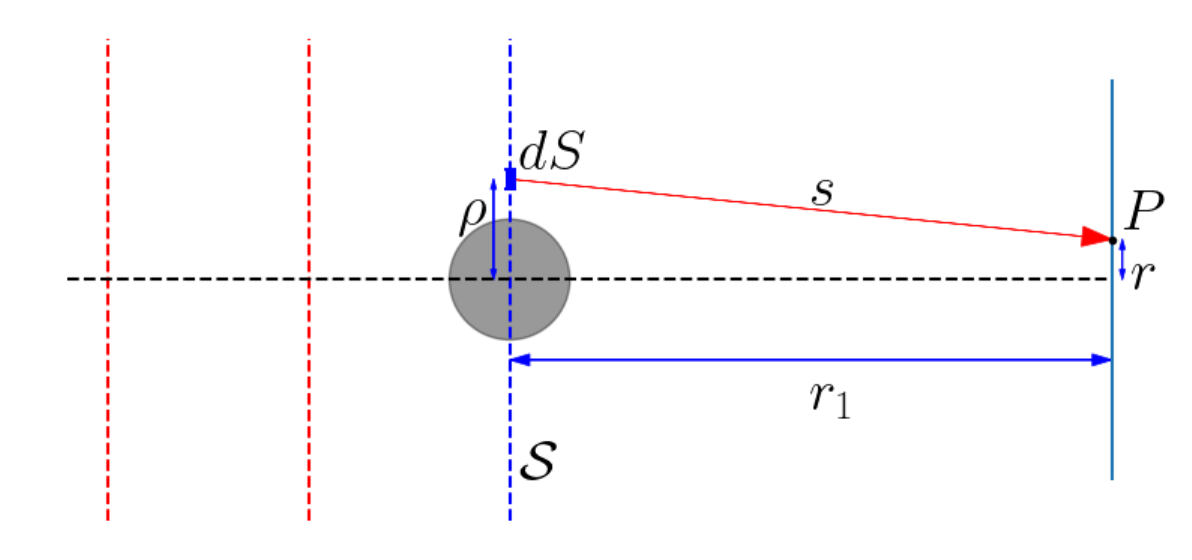}
    \caption{At large distances from the source, the incoming wavefronts from the left can be approximated by plane waves.  With this approximation, the incoming wave has the same amplitude and phase over the whole plane $\mathcal{S}$, simplifying the analysis of the interference pattern at the screen. The length of the displacement ${\bf s}$ is the distance from the surface element $dS$ (determined by radius ${\bf \rho}$) on plane $\mathcal{S}$ to the observation point $P$ on the screen.}
    \label{fig:integration-plane} 
\end{figure}

When the distance from the light source to the object is much larger than the size of the object, $r_0 \gg R$, the incoming light field upon reaching the object is to a good approximation a plane wave and the prefactor $A(S)=A_0$ is constant over the plane surface $\mathcal{S}$ shown in Figure~\ref{fig:integration-plane},
yielding the Huygens-Fresnel diffraction formula
\begin{equation}
U(P) = A_0 \int_\mathcal{S} \frac{e^{iks}}{s}dS,
\end{equation}
where the integration is taken over the surface $\mathcal{S}$ around the object.
To finally obtain the Huygens-Fresnel principle of secondary wavelets applied to the Poisson spot setting, one needs to assume an absorbing object, which in practice cuts away the cross section of the object from the plane $\mathcal{S}$. 

Due to cylindrical symmetry, the point $P$ can be chosen to lie in the plane of Fig.~\ref{fig:integration-plane} without loss of generality.
Expressing the surface integral in polar coordinates, the distance $s$ from the surface element $dS$ at distance $\rho$ from the center of the object and at the azimuthal angle $\theta$ is
\begin{eqnarray}
s &= \sqrt{r_1^2 + (r - \rho \sin \theta)^2 + (\rho \cos \theta)^2} \\
&= \sqrt{r_1^2 + r^2 - 2r\rho \sin \theta + \rho^2},
\end{eqnarray}
where $r$ is the distance of the observation point $P$ from the center of the shadow. 
The displacement field becomes now
\begin{equation}
U(r) = A_0 \int_{R}^\infty d\rho \int_0^{2\pi} d\theta\, \frac{\rho e^{ik\sqrt{r_1^2 + r^2 - 2r\rho \sin \theta + \rho^2}}}{\sqrt{r_1^2 + r^2 - 2r\rho \sin \theta + \rho^2}}.
\end{equation}

Notice that the phase factor in the integrand oscillates ever more rapidly for increasing distance $\rho$ from the center of the object.
This results in a cancellation of contributions far from the object, 
which in practice means that incoming plane waves need to be have constant amplitude and phase only in the vicinity of the object in order to satisfy the plane wave approximation.

Linearizing the square root yields
\begin{equation}
\sqrt{r_1^2 + r^2 - 2r\rho \sin \theta + \rho^2} \approx r_1  + \frac{r^2 - 2r\rho \sin \theta + \rho^2}{2r_1}
\end{equation}
and the integral becomes
\begin{equation}
U(r) = \frac{A_0 e^{ikr_1}}{r_1} e^{ik\frac{r^2}{2r_1}}\int_{R}^\infty d\rho \int_0^{2\pi} d\theta\, \rho e^{ik\frac{\rho^2- 2r\rho \sin \theta}{2r_1}}.
\end{equation}
Writing the expression in front of the integral as $u_0(r)$ yields in terms of wavelength $\lambda = 2\pi/k$,
\begin{equation}
U(r) = u_0 \int_{R}^\infty d\rho\, \rho e^{i\pi\frac{\rho^2}{\lambda r_1}} \int_0^{2\pi} d\theta\, e^{-i\pi\frac{2r\rho \sin \theta}{\lambda r_1}}.
\end{equation}
The angular integral is the zeroth order Bessel function $2\pi J_0(\pi\frac{2r\rho}{\lambda r_1})$, giving the final integral formula
\begin{equation}
U(r) = u_0 \int_{R}^\infty d\rho\, \rho e^{i\pi\frac{\rho^2}{\lambda r_1}} J_0(\pi\frac{2r\rho}{\lambda r_1}).
\label{eq:final_withoutscaling}
\end{equation}

It is instructive to analyze this integral equation a bit further and to make connections with the simple model described above.
Changing variables to $x = \rho- R$ yields
\begin{equation}
U(r) = u_0 e^{i\pi \frac{R^2}{\lambda r_1}}\int_0^\infty dx\, (R+x) e^{i\pi\frac{2Rx + x^2}{\lambda r_1}} J_0(\pi\frac{2r(R+x)}{\lambda r_1}).
\label{eq:shifted}
\end{equation}
The oscillating phase factor in the integrand means that the secondary wavelets emerging from larger distances $x$ are no longer in-phase and the constructive interference is lost. The length scale for this is given by the phase factor in the exponential

\begin{equation}
\frac{2Rx + x^2}{\lambda r_1} = 1,
\label{eq:lengthscale}
\end{equation}
yielding 
\begin{equation}
 x = \sqrt{R^2 - \lambda r_1}-R.   
\end{equation}
For typical parameters $\lambda r_1 \ll R^2$ allowing one to linearize the equation, which yields $x = \frac{\lambda r_1}{2R}$.
This can be immediately identified as the condition for the upper limit of the surface roughness in Eq.~\eqref{eq:surfacerough} up to a factor $\frac{1}{2}$, which is perhaps not surprising, since both values are related to the variation in $\rho$ that results in destructive interference. This also validates the central argument in the simple model that one can limit oneself to analyzing only the light originating from a narrow ring around the rim of the cross section of the obstacle, as light from further away will be increasingly cancelled by the averaging of the phase. This will be further validated numerically below.

The $r$-dependence of the amplitude $U(r)$ gives the spot size. The Bessel function is a decaying function, and the length scale for the spot size is given by the argument in the Bessel function
\begin{equation}
\pi \frac{2rR}{\lambda r_1} = 1,
\end{equation}
which yields $r = \frac{\lambda r_1}{2\pi R}$. This can be identified as the spot size obtained using the simple model above in Eq.~\eqref{eq:spotsize}, except for an additional factor $2/\pi$.
For the aforementioned typical lecture demonstration setup ($\lambda = 630\,\mathrm{nm}$, $r_1 = 10\,\mathrm{m}$, $R = 0.5\,\mathrm{mm}$) this would give Poisson spot radius of $2.0\,\mathrm{mm}$, which agrees even better with the observed spot of radius $2\,\mathrm{mm}$ in Figure~\ref{fig:demo}a.

\begin{figure}[h]
    \centering
    \includegraphics[width=0.9\columnwidth]{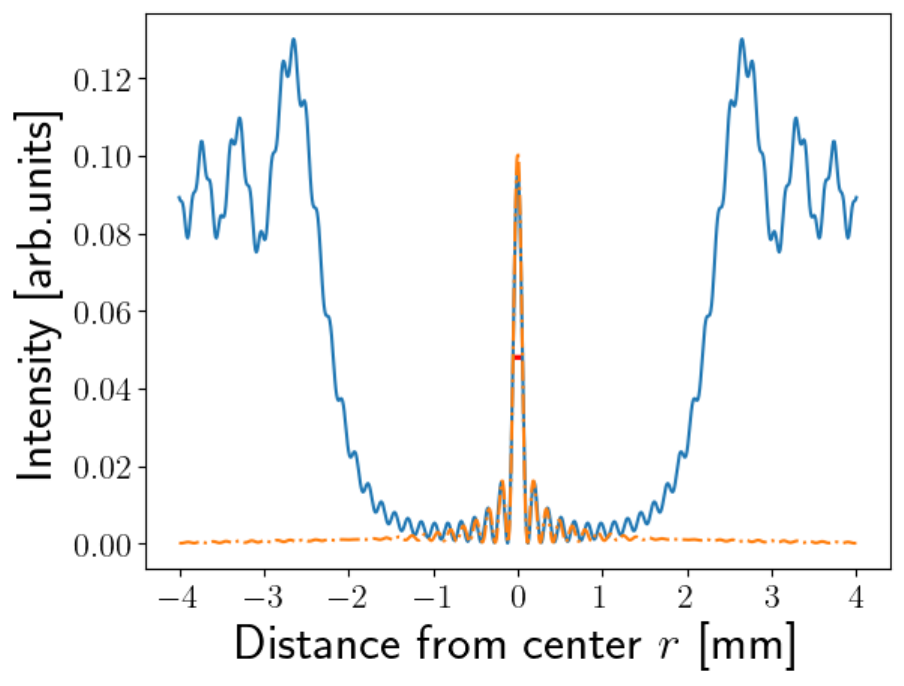}
    \caption{The intensity distribution $|U(r)|^2$ of the diffraction pattern as a function of the distance from the center (solid blue line) calculated using Eq.~\eqref{eq:converged}. The Poisson spot is visible as a sharp maximum in the center of a dark shadow. The parameters here are the wavelength $\lambda = 630\,\mathrm{nm}$, the object diameter $2R = 4\,\mathrm{mm}$ and distance to the screen $r_1=1\,\mathrm{m}.$ Shown is also the full-width at half-maximum of the central peak, with the width $\delta = 0.114\,\mathrm{mm}$. The orange dash-dot line shows the interference calculated using wavelets from only a very narrow ring around the object, as described in the text.}
    \label{fig:1dfigure} 
\end{figure}

Returning to Eq.~\eqref{eq:final_withoutscaling}, 
for numerical integration it helps to introduce a convergence factor $e^{-\rho^2/\sigma^2}$. While the $\rho$ dependence of the Bessel function $J_0$ guarantees convergence, the rapidly oscillating integrand is difficult to resolve in the numerical integration for large $\rho$. However, the oscillations mean that large values of $\rho$ contribute very little and the convergence factor acts as a soft cutoff for the integration.
One obtains
\begin{equation}
U(r) = u_0 \int_R^\infty d\rho\, \rho e^{i\pi\frac{\rho^2}{\lambda r_1}} e^{-\frac{\rho^2}{\sigma^2}} J_0(\pi\frac{2r\rho}{\lambda r_1}).
\label{eq:converged}
\end{equation}
The convergence factor can also describe a real physical effect, in which the intensity of the light decreases at large distances because of a possible intensity profile of the incoming light beam.
The convergence parameter used in all numerical results below is $\sigma = 0.015\,\mathrm{m}$, which corresponds to a two-centimeter wide Gaussian beam. This is already wide enough that the results in Fig.~\ref{fig:1dfigure} show no dependence on the value of $\sigma$ and indeed larger than the actual laser beam width used in the experiment.

The intensity of the observed diffraction pattern corresponds to the square of the amplitude $|U(r)|^2$. Figure~\ref{fig:1dfigure} shows a typical Poisson spot diffraction pattern calculated from Eq.~\eqref{eq:converged}, with the prominent narrow central peak of width $\delta = 0.114\,\mathrm{mm}$, which can be compared with the analytical result from above
\begin{equation}
2r = \frac{\lambda r_1}{\pi R} \approx 0.10\,\mathrm{mm},
\end{equation}
for the parameters in the figure.
The Poisson spot intensity maximum is roughly equal to the intensity outside the shadow region~\cite{reisinger2017}. 

The shadow region is at $r \in [-2\,\mathrm{mm},2\,\mathrm{mm}]$, corresponding to the shadow of a diameter $4\,\mathrm{mm}$ object. The edge of the shadow is not sharp and multiple diffraction fringes can be seen especially outside of the shadow region, with the widths of the fringes becoming narrower away from the shadow region. This is qualitatively very similar to the pattern observed experimentally in Fig.~\ref{fig:demo}b, despite the experimental setup involving an expanding beam, which makes the overall shadow region increasing in size with the distance to the screen. On the other hand, notice that the increase in size due to the expanding beam involves only the shadow region, scaling from $4\,\mathrm{mm}$ to roughly $10\,\mathrm{cm}$ in the experimental setup, and not the central peak, which expands only due to the diffraction according to the result obtained in Eq.~\eqref{eq:spotsize}.

Figure~\ref{fig:1dfigure} shows also the diffraction pattern calculated from Eq.~\eqref{eq:shifted} but with the integration regime limited only to a narrow ring, up to the length scale $x = \lambda r_1/(2R)$. The figure shows that this limited integration region still captures very well the Poisson spot and even the adjacent fringes, validating the key assumptions made in the simple analysis of Section~\ref{sec:orders}. However, the model fails to describe the wider region and in particular the edge of the shadow region and beyond.

\begin{figure}[h]
    \centering
    \includegraphics[width=0.9\columnwidth]{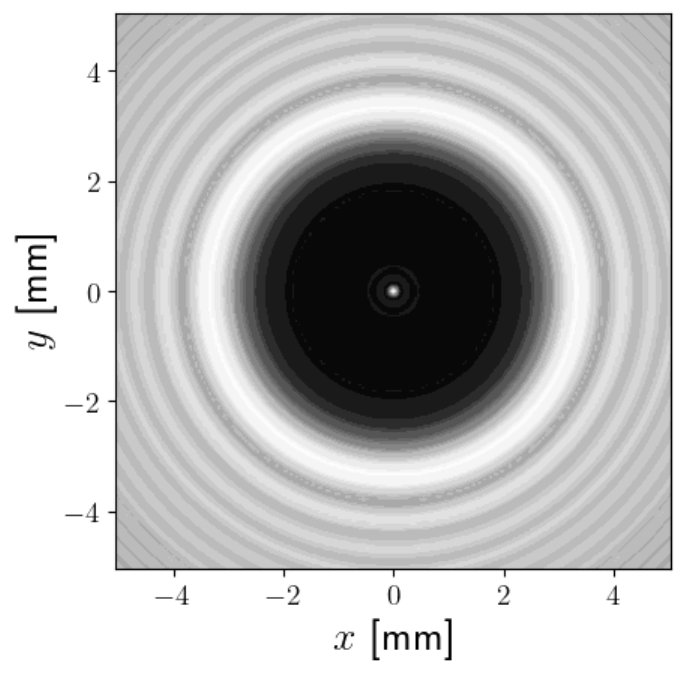}
    \caption{The two-dimensional intensity distribution $|U(x,y)|^2$ of the diffraction pattern. The Poisson spot is visible as a sharp maximum in the center of a dark shadow. The parameters here are the same as in Fig.~\ref{fig:1dfigure}.}
    \label{fig:2dfigure} 
\end{figure}

Finally, Figure~\ref{fig:2dfigure} shows the calculated two-dimensional diffraction pattern. The Poisson spot at the center and the interference fringes outside the shadow region are quite clear.

\section{Discussion}

The Poisson spot will create interest as a lecture demonstration, especially after posing the question in the title.
Moreover, it can be explained reasonably well with a simple model that builds on the familiar Huygens' principle that is often used in undergraduate teaching of the double slit experiment. 
The fact that the simple model requires some assumptions that are hard to justify {\it a priori} can also be seen as an advantage, as it is important for students to learn question the assumptions and models. Numerical and analytical calculations show, however, that the model can describe the characteristics of the Poisson spot very well. 

Finally, I believe that the mathematical formulation of the Huygens' principle in the Fresnel-Huygens diffraction formula is not necessarily too difficult for dedicated undergraduate students to understand. Indeed, we studied the Poisson spot in a planetary context in an April Fool's day project with first-year physics students~\cite{AprilFools}. That manuscript may be used as a light-hearted extra material when discussing diffraction and Poisson spot in undergraduate courses.

\begin{acknowledgements}
I would like to thank V. Havu for the question in the title and acknowledge wonderful discussions with students H. Viitasaari, O. Färdig, J. H. Siljander, A. P. Väisänen, A. S. Harju, and A. V. Nurminen.
\end{acknowledgements}

The author has no conflicts to disclose.

\end{document}